\documentclass[superscriptaddress,aps,prl,english,floatfix,twocolumn,10pt]{revtex4-2}
\usepackage[english]{babel}
\usepackage{graphicx}
\usepackage{float}
\usepackage{natbib}
\usepackage{amssymb}
\usepackage{amsmath}
\usepackage{mathtools}
\usepackage[utf8]{inputenc}
\usepackage{braket}
\usepackage{subfigure}
\usepackage{pgfplots}
\usepackage{csquotes}
\usepackage{hhline}
\usepackage{amssymb}
\usepackage{xr}

\usepackage{dcolumn}
\usepackage{tabularx}
\usepackage[colorlinks=true,linkcolor=blue,citecolor=blue,urlcolor=blue]{hyperref}
\usepackage{longtable}
\usepackage{listings}
\usepackage{color}
\usepackage[normalem]{ulem} 
\usepackage[export]{adjustbox}
\newcolumntype{C}{>{\centering\arraybackslash}X}

\begin{document}
\title{Topological boundaries in non-Hermitian p-wave Kitaev chains with Rashba spin-orbit coupling}

\author{Shahroze Shahab}
\affiliation{%
	Department of Physics and Astronomy, National Institute of Technology, Rourkela, Odisha-769008, India
}
\author{Aditi Chakrabarty}%
\email{aditichakrabarty030@gmail.com}
\affiliation{%
	Department of Physics and Astronomy, National Institute of Technology, Rourkela, Odisha-769008, India
}
\author{Sanjoy Datta}%
\email{dattas@nitrkl.ac.in}
\affiliation{%
	Department of Physics and Astronomy, National Institute of Technology, Rourkela, Odisha-769008, India
}
\date{\today}

\begin{abstract}

    In this work, we investigate the combined effects of Rashba spin-orbit coupling (RSOC) and non-Hermiticity on topological phase transitions in spinful p-wave Kitaev chains. While previous studies have separately examined non-Hermitian (NH) extensions of Kitaev chains and the effects of RSOC in Hermitian systems, the interplay between these two mechanisms remains largely unexplored. We analyze this interplay by considering two distinct types of complex on-site potentials: (i) a uniform gain/loss term and (ii) a complex quasiperiodic potential. We demonstrate that the impact of RSOC is highly model-dependent. In particular, RSOC does not affect the topological phase boundary in the Hermitian limit of the uniform gain/loss model (provided the spin-flip hopping is weaker than the pairing strength), but significantly alters the topological landscape in the NH regime. In contrast, for the quasiperiodic model, RSOC modifies the phase boundaries in both the Hermitian and non-Hermitian cases. Notably, we find that the combined interplay of non-Hermiticity and RSOC drives topological transitions at significantly lower potential strengths compared to the Hermitian limit. We derive analytical expressions for the topological phase transitions in both cases and validate our predictions through numerical calculations of energy spectra and real-space winding numbers. This work provides a comprehensive understanding of how non-Hermiticity and RSOC cooperatively reshape topological phase diagrams in one-dimensional superconducting systems.
\end{abstract}

\maketitle


\section{\label{sec:level1}INTRODUCTION}
Understanding the topological phases in superconductors began with the early works on the paradigmatic one-dimensional Kitaev chain\cite{Kitaev_2001}. A hallmark of this exactly solvable model is the emergence of Majorana zero modes (MZMs) under the open boundary condition (OBC) that appear at the two boundaries of the one-dimensional chain in the topologically non-trivial regime estimated using the periodic boundary condition (PBC). These MZMs are of significant interest for quantum computing applications due to their non-local nature, making them inherently protected against local perturbations and potentially useful for fault-tolerant quantum computation \cite{nayak2008nonQuantcomp,sarma2015majoranaQuantcomp,alicea2011nonQuantcomp}.\\
\indent
While the original Kitaev model describes spinless fermions with p-wave pairing, realistic implementations must account for spin-orbit coupling effects that are ubiquitous in materials of experimental interest. The interplay between superconductivity and Rashba spin-orbit coupling (RSOC) is particularly relevant for experimental platforms such as semiconductor nanowires \cite{birkholz2008spin} with proximity-induced superconductivity \cite{lutchyn2010majorana}, where RSOC serves as a crucial ingredient for realizing topological superconducting phases. Besides, the RSOC introduces significant modifications to the electronic band structure, including momentum-dependent spin splitting and the lifting of spin degeneracies. These effects can fundamentally alter topological properties and the phase transitions. In one of the past works \cite{wong2012majorana}, the authors realized a DIII class topological superconductor supporting Majorana Kramers doublets due to RSOC in $d$-wave superconductors. More recently, Silva et al. \cite{silva2024hybridization} explored hybridization-induced triplet superconductivity with $S_z = 0$ in chains of spinful fermions, revealing that such systems can host edge states with non-trivial topology under an applied magnetic field.\\
\indent On the other hand, dissipation and gain/loss mechanisms are unavoidable in real experimental systems and may substantially influence the electronic properties. In this regard, there has been growing interest in non-Hermitian (NH) quantum systems, which provide effective descriptions of open quantum systems that experience environmental coupling. The non-Hermiticity is generally achieved by introducing non-reciprocal hopping or a complex potential in the system. Moreover, non-Hermitian quasiperiodic systems, which lie between the periodic and disordered limits, often realize non-Hermiticity through a complex phase in the onsite potential that describes gain or loss in such systems \cite{longhi2019topological}. These NH systems can be realized in atomic \cite{xu2017weylAtomic,lee2014heraldedAtomic,li2019observationAtomic,yamamoto2019theoryAtomic}, optical \cite{chen2017exceptionalOptical,ozawa2019topologicaloptical,ozawa2019topologicaloptical,weimann2017topologicallyOptical}, electronic \cite{helbig2020generalizedElectronic,hofmann2020reciprocalElectronic},
and mechanical \cite{zhou2020nonMech,brandenbourger2019nonMech,ghatak2020observationMech} systems.
Interestingly, the non-Hermiticity introduces phenomena with no Hermitian counterparts, such as complex energy spectra, exceptional points, and the breakdown of the bulk-boundary correspondence. Furthermore, notably, the topological classification in the NH systems extends beyond the standard ten-fold scheme of Hermitian systems to a richer 38-fold classification \cite{gong2018topological,kawabata2019symmetry}.\\
\indent Previous works on NH Kitaev chains have identified rich topological phenomena in a multitude of systems \cite{wang2015spontaneous,sayyad2023topological,li2024anomalous}. Furthermore, Zeng et al.~\cite{zeng2016non} and Jiang et al.~\cite{Jiang_2021} investigated the role of complex on-site potentials, revealing modified bulk-boundary correspondence and unique topological features in the NH versions of the Kitaev chain. Notably, despite these advances, these works primarily focused on spinless models, leaving the effects of RSOC in the NH systems on topological superconductivity largely unexplored. Keeping this in mind, we systematically investigate the effect of RSOC in a spinful p-wave superconducting Kitaev chain and analytically understand the topological phase boundaries.\\
\indent More specifically, in this work, we demonstrate significant findings on the interplay between RSOC and non-Hermiticity with triplet p-wave pairing. First, we establish that, unlike the Hermitian case where topological phase boundaries remain fixed regardless of the RSOC strength, non-Hermiticity introduces an explicit dependence on the spin-flip hopping amplitude, shrinking the topological region with increasing strength of RSOC. To do so, we derive an analytical expression for these modified phase boundaries, validated using numerical estimations. In addition, for quasiperiodic potentials of non-Hermitian type, we prove that the critical potential strength for topological-trivial transition is altered, revealing how non-Hermiticity and RSOC cooperatively reshape the topological phase diagram. Finally, we identify regions where RSOC induces band crossings and gapless phases with emergent Fermi points that always appear in pairs.\\
\indent The remaining article is organized as follows: In Sec.~\ref{IIA}, we introduce the model Hamiltonian for the spinful p-wave Kitaev chain with RSOC. Secs.~\ref{IIB} and \ref{IIC} provide the real-space and momentum-space representations, respectively. 
In Section \ref{IV}, we analyze the NH superconductor, and discuss the details on the estimation of topological phase boundaries numerically. In Sec.~\ref{IV_new} we discuss the results of the two NH superconductors specifically considered in this work in detail. In particular, we provide analytical derivations of topological phase transitions of the NH system with gain/loss in Sec.~\ref{IVA} along with its numerical validations in Sec.~\ref{IVB}. In Section \ref{V}, we incorporate quasiperiodic potentials possessing non-Hermiticity, and demonstrate the findings using transfer matrix formalism in Sec.~\ref{VA} and Lyapunov exponent calculations in Sec.~\ref{VB} respectively, to analytically determine the topological regimes, along with the numerical verifications in Sec.~\ref{VC}. In the end, we summarize the crucial findings in Sec.~\ref{Conclusions}.

\section{RSOC in the spinful p-Wave Kitaev chain}
\subsection{Model Hamiltonian}
\label{IIA}
Here, we investigate the effect of RSOC \cite{birkholz2008spin} in one-dimensional spinful p-wave Kitaev chains \cite{Kitaev_2001} with complex on-site potentials. Similar Kitaev chains have been recently gaining attention in several important contexts \textcolor{blue}{\cite{zeng2016non,wang2015spontaneous,Jiang_2021}}. The Hamiltonian is defined as,
\begin{equation}
	\begin{split}
		\mathcal{H} & = J \sum_{n,\sigma}
		\left(c_{n+1,\sigma}^{\dagger} c_{n,\sigma} +  \text{H.c.}\right)+ \Delta \sum_{n,\sigma,\sigma'}
		\left( c_{n+1,\sigma}^{\dagger} c_{n,\sigma'}^{\dagger}
		+ \text{H.c.} \right)                            \\
		            & - \alpha_z \sum_{n,\sigma,\sigma'}
		\left( c_{n+1,\sigma}^{\dagger} (i\sigma_y)_{\sigma,\sigma'} c_{n,\sigma'} + \text{H.c.} \right) + \sum_{n,\sigma} V_n' c_{n,\sigma}^{\dagger} c_{n,\sigma}.
	\end{split}
	\label{eq:hamiltonian}
\end{equation}

Here, $c_{n,\sigma}^{\dagger}$ and $c_{n,\sigma}$ are the creation and annihilation operators for the fermion at site $n$ with spin $\sigma,\sigma'(\uparrow,\downarrow)$. We consider $\sigma \neq \sigma'$ in our work. $J$ denotes the fermionic hopping amplitude, $\Delta$ (which is taken to be real) is the term which brings in the p-wave superconducting pairing between fermions of antiparallel spins in neighboring sites of the wire and $\alpha_z$ is the spin-flip amplitude of the fermion due to RSOC. $\sigma_y$ is the y-component of Pauli's spin matrices, and $V_n'$ symbolizes the complex on-site potential which induces non-Hermiticity. We consider two cases of the on-site potential containing non-Hermiticity, which will be defined later in the manuscript.

\subsection{Real-space representation}
\label{IIB}
To obtain the representation of the 1D spinful superconducting chain in the real space, we define the Nambu spinor $\Psi_{n}=(c_{n,\uparrow},c_{n,\uparrow}^\dagger,c_{n,\downarrow},c_{n,\downarrow}^{\dagger})^{T}$. Using the BdG formalism, the Hamiltonian in real space is then given as,
\begin{equation}
	\mathcal{H}_{BdG} = \frac{1}{2}\Psi^\dagger H_{BdG} \Psi  + \text{constant,}
	\label{eq:bdg}
\end{equation}
where

\begin{equation}
	\begin{aligned}
		H_{BdG} = & \sum_{n} J \left( |n\rangle \langle n+1|
		+ |n+1\rangle \langle n| \right) \otimes \sigma_0 \otimes \sigma_z                                                                     \\
		              & + \Delta \sum_{n} \left( |n\rangle \langle n+1| - |n+1\rangle \langle n| \right) \otimes \sigma_x \otimes i\sigma_y    \\
		              & + \alpha_z \sum_{n} \left( |n\rangle \langle n+1| - |n+1\rangle \langle n| \right) \otimes i \sigma_y \otimes \sigma_z \\
		              & + \sum_{n} V_n' |n\rangle \langle n| \otimes \sigma_0 \otimes \sigma_z.
	\end{aligned}
	\label{eq:bdg_hamiltonian}
\end{equation}
In the above equation, $\sigma_0$ is the identity matrix while $\sigma_{x,y,z}$ are the usual components of Pauli spinors. $\Psi = (\Psi_1, \Psi_2, \Psi_3, \dots, \Psi_{N-1}, \Psi_N)^T$, where $N$ represents the number of sites in the chain. We use this real-space form of the Hamiltonian to numerically obtain the energy spectrum and estimate the topological winding number that will be presented in the subsequent discussions.

\subsection{$k$-space representation}
\label{IIC}
For an infinite chain with periodic boundaries, one can Fourier transform the Hamiltonian in Eq.~\eqref{eq:hamiltonian} to obtain the Hamiltonian in the momentum space using the transformation $c_{n,\sigma(\sigma')} = \frac{1}{\sqrt{N}} \sum_{k} e^{i k n} c_{k,\sigma(\sigma')}$. We introduce the $k$-space spinor as $\Psi_{k} = (c_{k,\uparrow},c_{-k,\uparrow}^{\dagger},c_{k,\downarrow},c_{-k,\downarrow}^{\dagger})^{T}$. The transformed Hamiltonian is then retrieved as,
\begin{equation}
	\begin{aligned}
		\mathcal{H}(k) = 2J \cos(k) c_{k, \sigma}^\dagger c_{k, \sigma} + 2i \Delta \sin(k) \left( c_{-k,\uparrow}^\dagger c_{k,\downarrow}^\dagger - c_{k,\downarrow} c_{-k,\uparrow} \right) \\
		+ 2i \alpha_z \sin(k) \left( c_{k,\uparrow}^\dagger c_{k,\downarrow} - c_{k,\downarrow}^\dagger c_{k,\uparrow} \right)
		+ \sum_{k,\sigma} V' c_{k, \sigma}^\dagger c_{k, \sigma}. ~~~~~~~
	\end{aligned}
	\label{eq:bdg_hamiltonian_momentum}
\end{equation}
Similar to Eq.~\eqref{eq:bdg}, we can write,
\begin{equation}
	\mathcal{H}(k)_{BdG} = \frac{1}{2}\sum_{k} \Psi_{k}^{\dagger} H{(k)}_{BdG} \Psi_{k} + \text{constant.}
\end{equation}
Here, $H(k)_{BdG}$ is given by
\small
\begin{equation}
	{H}(k)_{BdG} = \begin{pmatrix}
		\xi_k                       & 0                          & 2 i\alpha_z \text{sin}(k) & -2 i \Delta \text{sin}(k)   \\
		0                           & -\xi_k                     & 2 i \Delta \text{sin}(k)  & -2 i \alpha_z \text{sin}(k) \\
		-2 i \alpha_z \text{sin}(k) & -2 i \Delta \text{sin}(k)  & \xi_k                     & 0                           \\
		2 i \Delta \text{sin}(k)    & 2 i \alpha_z \text{sin}(k) & 0                         & -\xi_k
	\end{pmatrix}
	\label{eq:bdg_hamiltonian_momentum_matrix}
\end{equation}
\normalsize
where $\xi_k = 2J\cos(k) + V'$. A detailed derivation of the Hamiltonian in momentum space is presented in Appendix \ref{App_A}.

\section{Topology in non-Hermitian Kitaev chains with RSOC}
\label{IV}
In this work, as previously mentioned, we consider the non-Hermiticity induced by the complex on-site potential similar to that of Refs.~\cite{zeng2016non} and \cite{Jiang_2021}. Specifically, we consider the two well-studied cases in literature, i.e., a potential with an on-site loss/gain and the NH version of the quasiperiodic potential as analyzed in Ref.~\cite{longhi2019topological}. Our aim is to deeply scrutinize the effect of the RSOC on the topological transition numerically as well as by providing analytical arguments in both these models one by one in the following discussion. Numerically, we use the exact diagonalization on the BdG Hamiltonian given in Eq.~\eqref{eq:bdg_hamiltonian} throughout the manuscript. Henceforth, we fix $J = 1$ as the energy scale, and use the p-wave superconducting pairing $\Delta = 0.5$. Although we fix \(\Delta=0.5\) for the numerics, the results are qualitatively robust. Increasing \(\Delta\) enlarges the bulk gap and increases the topological regime, while decreasing \(\Delta\) narrows it and eventually destroys the topology. The qualitative features depend only the competition between $\alpha_z$ and $\Delta$. In the entire work, we have considered that the Kitaev chain consists of $100$ lattice sites, and considered OBC and PBC for numerical determination of the energy band diagrams and the real space winding number (that is introduced in the following discussion), respectively.

\subsection{Numerical recipe: the real space winding number}
One of the main aims in this work is to numerically figure out the topological phase boundaries in both the non-Hermitian $p$-wave Kitaev chains given by the Hamiltonian $\mathcal{H}$, as already discussed. It is important to note that $H_{BdG}$ possesses sublattice symmetry, i.e. $\Gamma H_{BdG} \Gamma^{-1} = -H_{BdG}$ where $\Gamma$ is the sublattice symmetry operator and is given as $\Gamma =  \mathbf{1} \otimes \sigma_x$. 

For such Hamiltonians, it is well known that the winding number of $H_{BdG}$ can be deduced from the knowledge of the winding number of the matrix $Q$ which is obtained using the polar decomposition wherein any invertible matrix can be written as $H_{BdG} =  QP$. Here, $Q$ is the unique unitary matrix and $P$ is a positive semi-definite Hermitian matrix. Using this, one can easily show that the Hamiltonian $H_{BdG}$ is homotopic to $Q$ (please refer to Theorem 1, Section IV in Ref.~\cite{gong2018topological}). This also implies that $H_{BdG}$ can be deformed continuously to $Q$ without closing the energy gap, leading to the formulation of the real space winding number in terms of the matrix Q defined as $W = \frac{1}{L'} \mathrm{Tr}(\Gamma Q^{\dagger}[Q, X])$ \cite{roy2023critical,claes2021skin,kroy2024quasiperiodic}, where X is the position operator. The trace is taken over the bulk of the system with the given length $L^\prime$ = N/2 to reduce edge effects.

\section{Results and Discussions}\label{IV_new}
\subsection{Case-i: $V_n' = \mu + i \gamma$}
\label{IVA_new}
\subsubsection{Analytical understanding of the topological phase transition}\label{IVA}

	In this section, we follow the approach used in Ref.~\cite{yokomizo2019non} in the NH system with a complex potential and obtain the generalized Bloch Hamiltonian from Eq.~\eqref{eq:bdg_hamiltonian_momentum_matrix} using the relation $\beta = e^{ik}$ as,
	\begin{equation}
		{H}(\beta) =
		\begin{pmatrix}
			\xi_\beta              & 0                     & \alpha_z \delta_\beta & -\Delta \delta_\beta   \\
			0                      & -\xi_\beta            & \Delta \delta_\beta   & -\alpha_z \delta_\beta \\
			-\alpha_z \delta_\beta & -\Delta \delta_\beta  & \xi_\beta             & 0                      \\
			\Delta \delta_\beta    & \alpha_z \delta_\beta & 0                     & -\xi_\beta
		\end{pmatrix},
	\end{equation}

	where
	\begin{equation}
		\xi_\beta = J (\beta + \beta^{-1}) + V_n', \text{ and}\quad
		\delta_\beta = (\beta - \beta^{-1}).
	\end{equation}

	The 4$\times$4 Bloch Hamiltonian gives rise to a 4-band energy spectrum where the eigenvalues are given by $\epsilon_{1,\pm}(\beta)$ and $\epsilon_{2,\pm}(\beta)$, and are obtained as,
  \begin{eqnarray}
    \epsilon_{1\pm}(\beta) = \pm \sqrt{-\delta_{\beta}^2 (\alpha_z^2+\Delta^2) + \xi_{\beta}^2 + 2 \sqrt{\alpha_z^2 \delta_{\beta}^2 (\delta_{\beta}^2 \Delta^2-\xi_{\beta}^2)}}; \nonumber\\
    \epsilon_{2\pm}(\beta) = \pm \sqrt{-\delta_{\beta}^2 (\alpha_z^2+\Delta^2) + \xi_{\beta}^2 - 2 \sqrt{\alpha_z^2 \delta_{\beta}^2 (\delta_{\beta}^2 \Delta^2-\xi_{\beta}^2)}}. \nonumber\\
    \label{eq:eigenvalues}
  \end{eqnarray}

	In addition, the condition $\text{det}(\mathcal{H}_{\beta}) = 0$ signals the presence of at least one localized zero energy mode. This gives rise to,

	\begin{equation}
		(\delta_{\beta}^2 (\alpha_z^2 - \Delta^2) + \xi_{\beta}^2)^2 = 0.
    \label{eq:det}
	\end{equation}

	On substituting the value of $\xi_{\beta}$ from Eq.~\ref{eq:det} into Eq.~\ref{eq:eigenvalues}, $\epsilon_{1,\pm}(\beta)$ turns out to be zero. From here on, we denote the eigenstates corresponding to these two zero-energy modes by the subscript '$\pm$'.

  Furthermore, Eq.~\ref{eq:det}, which is a polynomial in $\beta$, has 4 roots which we denote by $\beta_{1\pm}$ and $\beta_{2\pm}$ and are given as,

	\begin{equation}
		\beta_{1\pm} = \frac{-V_n' \pm \sqrt{{V_n'}^2 - 4\bigl(J^2 - {\bar{\Delta}}^2\bigr)}}{2(J + {\bar{\Delta}})},
	\end{equation}

	\begin{equation}
		\beta_{2\pm} = \frac{-V_n' \pm \sqrt{{V_n'}^2 - 4\bigl(J^2 - {\bar{\Delta}}^2\bigr)}}{2(J - {\bar{\Delta}})}.
	\end{equation}

	where $\bar{\Delta} = \sqrt{\Delta^2 - \alpha_z^2}$.

	Following Ref.~\cite{doi:10.7566/JPSJ.91.124711}, we find the boundary zero modes under the OBC in the presence of RSOC. For large $N$, the right eigenvector of the real-space Hamiltonian $\mathcal{H}$ from Eq.~\ref{eq:bdg_hamiltonian} with zero eigenvalues can be formed by a linear superposition of the solutions written as,

	\begin{equation}
		|\phi^{R}\rangle =
		\sum_{j=1}^{N} |j\rangle \,\Bigl[\,
			c_{+}\,\beta_{+}^{\,j}
			\begin{pmatrix}
				u_{+,\uparrow}   \\
				v_{+,\uparrow}   \\
				u_{+,\downarrow} \\
				v_{+,\downarrow}
			\end{pmatrix}
			\;+\;
			c_{-}\,\beta_{-}^{\,j}
			\begin{pmatrix}
				u_{-,\uparrow}   \\
				v_{-,\uparrow}   \\
				u_{-,\downarrow} \\
				v_{-,\downarrow}
			\end{pmatrix}
			\Bigr].
	\end{equation}

	where $c_{\pm}$ are arbitrary constants and $|j\rangle$ denotes the eigenvector at $j$th site.  The imposition of OBC indicates that the coefficient of $|j \rangle$ vanishes at $j=0$ and $j=N$. This is satisfied if $|\beta_{+}|<1$ and $|\beta_{-}|<1$ in addition to
	\begin{equation}
		c_{+}
		\begin{pmatrix}
			u_{+,\uparrow}   \\
			v_{+,\uparrow}   \\
			u_{+,\downarrow} \\
			v_{+,\downarrow}
		\end{pmatrix}
		\;+\;
		c_{-}
		\begin{pmatrix}
			u_{-,\uparrow}   \\
			v_{-,\uparrow}   \\
			u_{-,\downarrow} \\
			v_{-,\downarrow}
		\end{pmatrix} =0.
	\end{equation}

	The above solution indicates the presence of an eigenvector localized near the left end of the chain. Similarly, one can construct an eigenvector localized near the right end by imposing the condition $|\beta_{+}|>1$ and $|\beta_{-}|>1$ along with the condition:

	\begin{equation}
		c_{+}\,\beta_{+}^{\,N}
		\begin{pmatrix}
			u_{+,\uparrow}   \\
			v_{+,\uparrow}   \\
			u_{+,\downarrow} \\
			v_{+,\downarrow}
		\end{pmatrix}
		\;+\;
		c_{-}\,\beta_{-}^{\,N}
		\begin{pmatrix}
			u_{-,\uparrow}   \\
			v_{-,\uparrow}   \\
			u_{-,\downarrow} \\
			v_{-,\downarrow}
		\end{pmatrix}=0.
	\end{equation}

	In addition, since $\beta_{2 \pm} = \beta_{1 \mp}^{-1}$, the condition for right localization, $|\beta_{1,+}|>1$ and $|\beta_{1,-}|>1$ implies the simultaneous existence of another left localized mode as $|\beta_{2,+}|<1$ and $|\beta_{2,-}|<1$. These conditions yield,

	\begin{equation}
		|\mu| < 2J \sqrt{1 - \left( \frac{\gamma}{2\bar{\Delta}} \right)^2}.
		\label{eq:phase_transition_non_hermitian}
	\end{equation}
	 Interestingly, our analytical estimate yields an effective superconducting pairing strength $\bar{\Delta}$, which is dependent upon the strength of RSOC, i.e., $\alpha_z$. To derive an analytical expression for the phase boundary for the topological transition, we will use the above equation, which yields,
\begin{eqnarray}
\frac{|\mu|^2}{(2J)^2} &<& 1 - \left(\frac{\gamma}{2\bar{\Delta}}\right)^2 \nonumber \\
\implies \frac{|\mu|^2}{(2J)^2} + \left(\frac{\gamma}{2\bar{\Delta}}\right)^2 &<& 1 \nonumber \\
\implies \frac{\mu^2}{(2J)^2} + \frac{\gamma^2}{(2\bar{\Delta})^2} &<& 1
\end{eqnarray}

The phase boundary which is an ellipse is then given by the equation $\frac{\mu^2}{4J^2} + \frac{\gamma^2}{4\bar{\Delta}^2} = 1$. In the following discussion, we will verify this topological transition numerically and investigate the effect of RSOC from the topological aspects.

\subsubsection{Numerical determination of $\mu_t$}
\label{IVB}


\begin{figure}[ht!]
  \begin{tabular}{p{\linewidth}c}
      \centering
      \includegraphics[width=0.48\columnwidth, height=3.6cm]{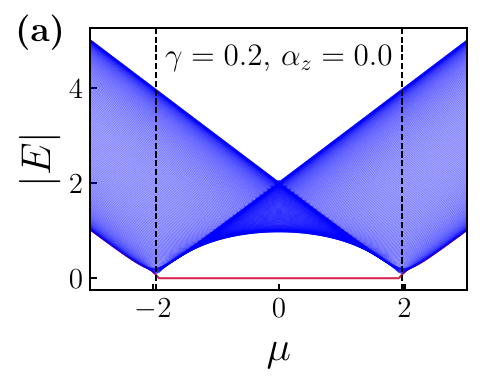}
      \includegraphics[width=0.48\columnwidth, height=3.6cm]{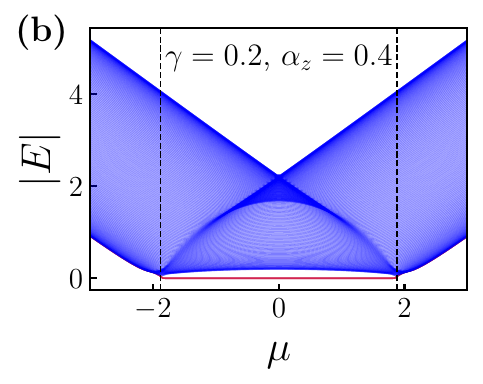}\\
      \caption{$|E|$ as a function of the real part $\mu$ of the NH on-site potential with $\gamma=0.2$ under OBC at (a) $\alpha_z=0.0$ (without the RSO interaction) and (b) $\alpha_z = 0.4$. The red points indicate the presence of MZMs while the vertical dashed lines indicate the analytically calculated transition point,i.e., $|\mu| = 2J \sqrt{1 - \left( {\gamma}/{2\bar{\Delta}} \right)^2}$.}
      \label{fig:abs_E_vs_mu_NH}
  \end{tabular}
\end{figure}

\begin{figure*}[t!] 
  \begin{tabular}{p{\linewidth}c}
      \centering
      \includegraphics[width=0.225\textwidth, height=3.8cm]{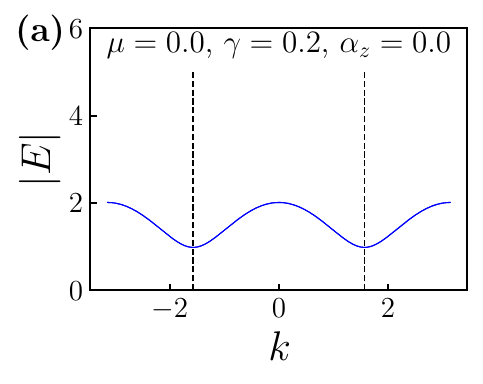}
      \includegraphics[width=0.225\textwidth, height=3.8cm]{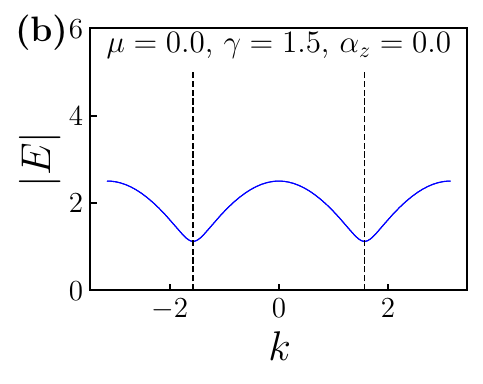}
      \includegraphics[width=0.225\textwidth, height=3.8cm]{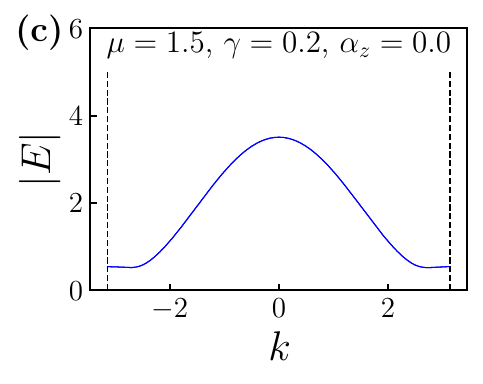}
      \includegraphics[width=0.225\textwidth, height=3.8cm]{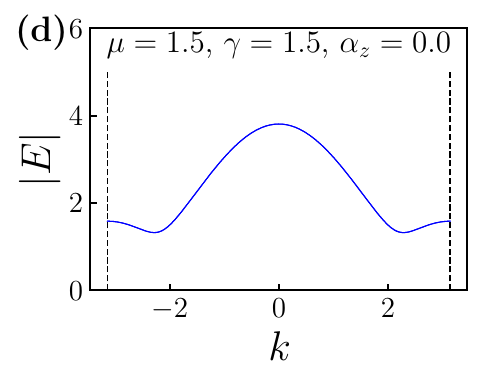}\\
      \includegraphics[width=0.225\textwidth, height=3.8cm]{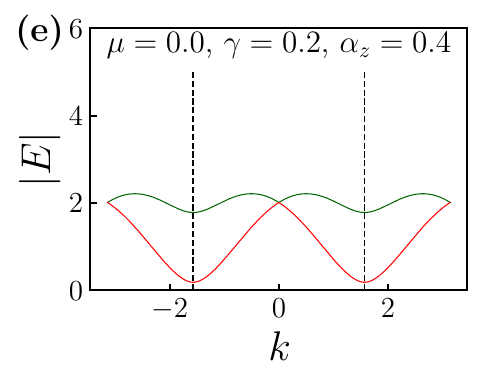}
      \includegraphics[width=0.225\textwidth, height=3.8cm]{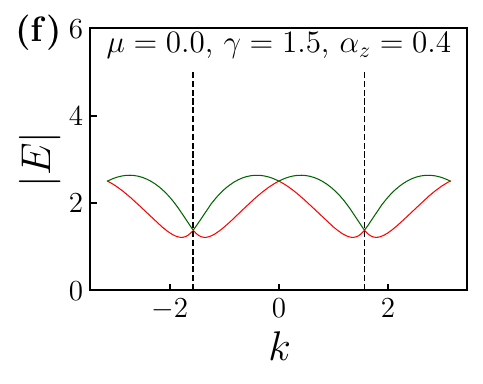}
      \includegraphics[width=0.225\textwidth, height=3.8cm]{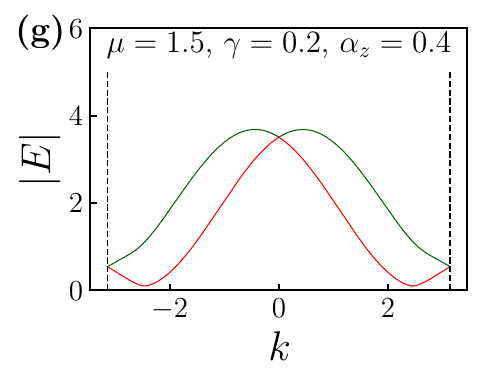}
      \includegraphics[width=0.225\textwidth, height=3.8cm]{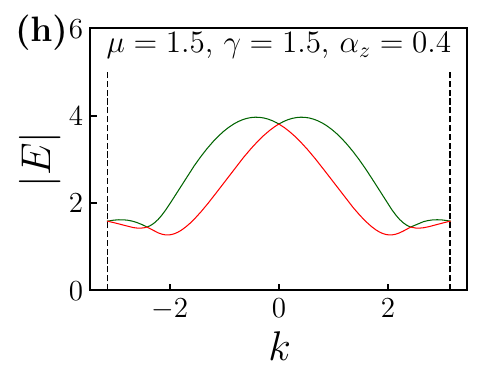}
      \caption{Absolute value $|E|$ of the energy eigenvalues as a function of momentum $k$ for various parameters $\alpha_z$, $\gamma$, and $\mu$. Panels (a)-(d) exhibit the case without RSOC ($\alpha_z = 0$), showing degenerate overlapping bands, while panels (e)-(h) demonstrate the effects of RSOC ($\alpha_z = 0.4$) with distinct band splitting. The energy bands derived in Eq.~\ref{eq:eigenvalues} are represented by different colors, i.e., $|\epsilon_1(\beta)|$ is represented in green and $|\epsilon_2(\beta)|$ is represented in red. Notably, even in the presence of non-Hermiticity, the bands maintain degeneracy at the time-reversal invariant momenta $k=0$ and $k=\pi$.}
      \label{fig:numerical_abs_spectrum}
  \end{tabular}
\end{figure*}

In this section we numerically determine the critical value of the NH on-site potential $\mu_t$ for topological phase transition. We plot the absolute values of the eigenenergies $|E|$ as a function of $\mu$ for two different values of $\alpha_z$ in Fig.~\ref{fig:abs_E_vs_mu_NH}. We find that the numerically derived phase transition point agrees well with the analytically predicted value. It is interesting to note that in stark contrast to the Hermitian case ($\gamma=0$) where $\mu_t$ is independent of $\alpha_z$ (as demonstrated in Appendix~\ref{App:App_B}), when the on-site potential is complex ($\gamma \neq 0$), the topological regime shrinks with an increase in $\alpha_z$. To have a clearer understanding on the effects of parameters $\mu$, $\gamma$ and $\alpha_z$ in the topologically non-trivial regime, we present the band structures from Eqs.~\ref{eq:eigenvalues} in Fig.~\ref{fig:numerical_abs_spectrum}. It is clear that in the absence of RSOC, $\epsilon_{1\pm}(\beta)$ and $\epsilon_{2\pm}(\beta)$ are identical, and all the four bands are degenerate (as shown in blue), which is clear from Figs.~\ref{fig:numerical_abs_spectrum}(a)-(d), irrespective of the values of $\mu$ and $\gamma$.  However, in the presence of RSOC ($\alpha_z=0.4$), $\epsilon_{1\pm}(\beta)$ and $\epsilon_{2\pm}(\beta)$ are non-identical, and the energy bands split as demonstrated in Figs.~\ref{fig:numerical_abs_spectrum}(e)-(h). In addition, when $\mu=0$, the energy bands show their minimum at $k=\pi/2$ and maximal values at $k=0$ and $k=\pi$ both in the presence and absence of $\alpha_z$ (as is evident from Figs.~\ref{fig:numerical_abs_spectrum}(a),(b),(e) and (f). This suggests the dominance of $cos^2$ term in the energy bands suggests that the hopping and the on-site potential (which is non-zero) dominates and determines the overall energy bands of the Hamiltonian. 
On the other hand, when $\mu=1.5$, the minimum at $k=\pi/2$ no longer exists. However, it is interesting to note that even in the NH case the bands $\epsilon_{1\pm}(\beta)$ and $\epsilon_{2\pm}(\beta)$ are still degenerate at the time-reversal invariant points $k=0$ and $k=\pi$, similar to what is remarked and observed in the past literature.

\begin{figure}[ht!]
  \begin{tabular}{p{\linewidth}c}
      \centering
      \includegraphics[width=0.48\columnwidth, height=3.6cm]{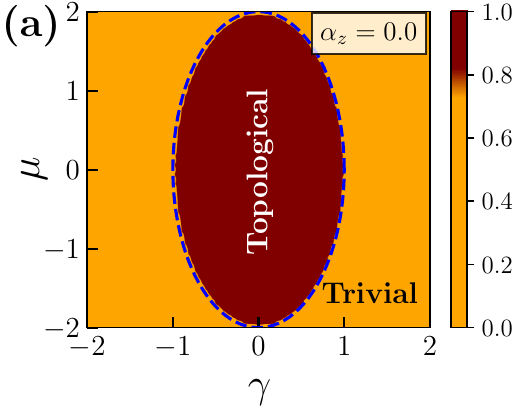}
      \includegraphics[width=0.48\columnwidth, height=3.6cm]{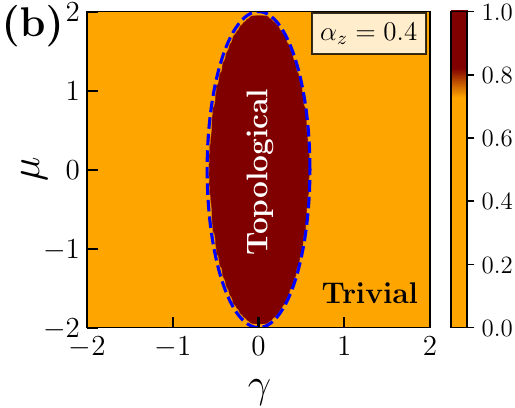}\\
      \caption{The real-space winding number of the system in the parameter space of the real ($\mu$) and the imaginary part ($\gamma$) of the NH potential at (a) $\alpha_z=0.0$ (without the RSOC) and (b) $\alpha_z = 0.4$(with RSOC). The maroon region (W = 1) indicates the topologically non-trivial regime, while the yellow region (W=0) represents the topologically trivial phase. The blue dashed lines indicate the analytical phase transition boundaries calculated using Eq.~\ref{eq:phase_transition_non_hermitian}.}
      \label{fig:fig5}
  \end{tabular}
\end{figure}

\begin{figure}[h!]
  \begin{tabular}{p{\linewidth}c}
      \centering
      \includegraphics[width=0.48\columnwidth, height=3.6cm]{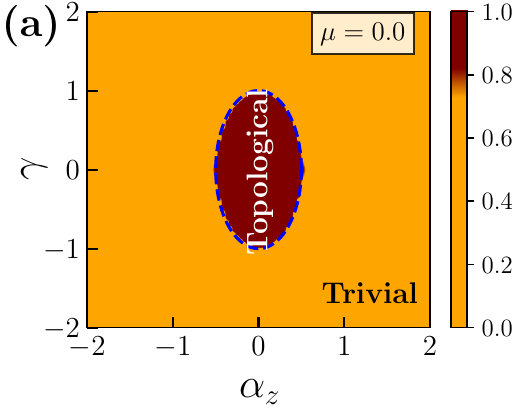}
      \includegraphics[width=0.48\columnwidth, height=3.6cm]{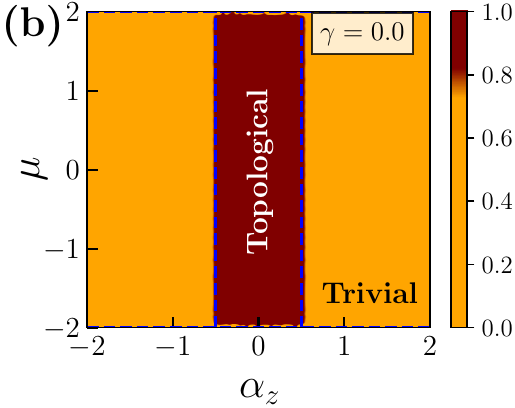}\\
      \includegraphics[width=0.48\columnwidth, height=3.6cm]{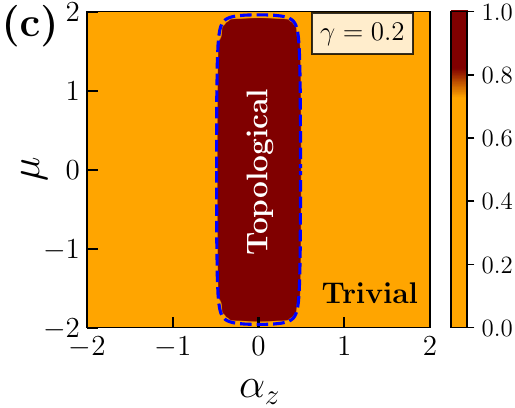}
      \includegraphics[width=0.48\columnwidth, height=3.6cm]{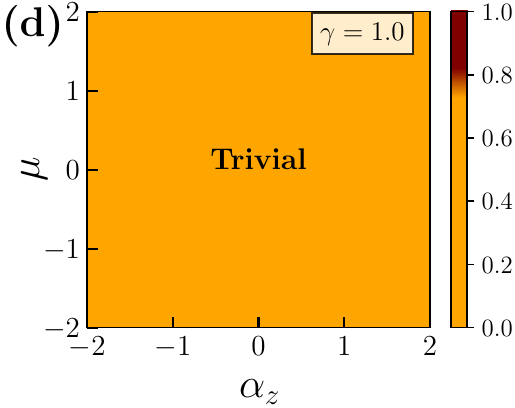}\\
      \caption{The real-space winding number of the system. (a) The phase diagram in the parameter space of $\gamma$ as a function of the strength of RSOC ($\alpha_z$). (b)-(d) The phase diagrams in the parameter space of the $\mu$ $vs.$ $\alpha_z$ for increasing values of the imaginary part $\gamma$. The blue dashed lines indicate the analytical phase transition boundaries, as already mentioned earlier.}
      \label{fig:fig6}
  \end{tabular}
\end{figure}

To complete our understanding on the combined effect of the parameters $\mu$, $\gamma$ and $\alpha_z$, we plot the numerical and analytical topological phase boundaries separating the trivial and non-trivial regimes in Figs.~\ref{fig:fig5}(a-b). From Fig.~\ref{fig:fig5}(b) is clear that the topological phase boundary shrinks when $\alpha_z\neq0$ as compared to the case when $\alpha_z=0$ as shown in Fig.~\ref{fig:fig5}(a). The numerical results agree excellently with our analytical understanding given in Eq.~\ref{eq:phase_transition_non_hermitian}. Interestingly, in Fig.~\ref{fig:fig6}(a), when the potential is completely complex, we obtain the phase diagram separating the topologically trivial and non-trivial phases, with contributions from both $\gamma$ and $\alpha_z$, in sharp contrast to the Hermitian case as discussed in Appendix~\ref{App:App_B}. In addition, from Figs.~\ref{fig:fig6}(b-d), it is evident that the topological regime shrinks with an increase in the gain/loss parameter $\gamma$ and vanishes when $\gamma=1$, irrespective of the strength of $\mu$ and $\alpha_z$. This is clear from Eq.~\ref{eq:phase_transition_non_hermitian} as $\mu$ becomes undefined when $\gamma > 2 \bar{\Delta}$ which is true when $\gamma=1$.

\subsection{Case-ii: $V'_n= V \text{cos}(2\pi\alpha n+ih)$}
\label{V}
Another dimension of complexity arises from quasiperiodic potentials, which lie between perfectly periodic and random potentials \cite{sokoloff1985unusual}.  These potentials have been implemented in various experimental systems, including optical lattices \cite{roati2008anderson}, photonic crystals \cite{lahini2009observationPhotonic,verbin2013observation}, cavity-polaritons \cite{tanese2014fractalCavity,goblot2020emergenceCavity}, and moiré structures \cite{balents2020superconductivityMoire}. Unlike periodic potentials that yield extended Bloch states or fully random potentials that cause Anderson localization in 1D even for arbitrary strength of random potential \cite{anderson1958absence}, quasiperiodic potentials can induce a localization phase transition and an associated topological transition in non-Hermitian systems \cite{longhi2019topological,quasiperiodicNHTapan,quasiperiodicNHTang,quasiperiodicNHJiang}.
In this section, we investigate the NH Kitaev chain with a quasiperiodic on-site potential that is incommensurate with the underlying lattice. The potential is given by $V_n' = V \cos(2\pi \alpha n + i h)$, where $V$ represents the amplitude, $\alpha$ is the inverse golden ratio ($\frac{\sqrt{5}-1}{2}$), and $h$ is the imaginary phase in the quasiperiodic potential that induces the non-Hermiticity in this case. This model encompasses several important limiting cases: when $h=\Delta=\alpha_z=0$, it reduces to the well-studied Hermitian Aubry-André-Harper model \cite{aubry1980}, which exhibits a topological phase transition at $V = 2J$, as discussed. Moreover, the scenario with $\Delta \neq 0$ but $h=\alpha_z=0$ has been previously investigated in the context of a Hermitian spinless Kitaev chain \cite{degottardi2013majorana}.
Building on our analysis of the uniform NH system in previous sections, we now extend these results by examining the combined effects of quasiperiodicity, non-Hermiticity, superconducting pairing, and RSOC. We demonstrate that these interplays lead to rich topological phase diagrams with transition boundaries that depend on both $\alpha_z$ and $h$, which will be discussed further.

\subsubsection{The transfer matrix method}
\label{VA}
To analyze the topological properties of such a non-Hermitian quasiperiodic Kitaev chain, we employ the transfer matrix formalism, which allows us to determine the localization properties of eigenstates. We first transform our Hamiltonian to the Majorana basis. The Dirac fermionic operator $c_{n,\sigma}$ can be expressed in terms of Majorana fermions, $c_{n,\sigma} = (a_{n,\sigma} + i b_{n,\sigma})/{2}$ where $a_{n,\sigma}$ and $b_{n,\sigma}$, which are Hermitian operators satisfying the anticommutation rules $\{a_{n,\sigma}, b_{m,\sigma'}\} = 0$ and $  \{a_{n,\sigma}, a_{m,\sigma'}\} =  \{b_{n,\sigma}, b_{m,\sigma'}\} =  2 \delta_{n,m} \delta_{\sigma,\sigma'}$. The Hamiltonian then becomes,
\begin{align}
	\mathcal{H} = \frac{i}{2} \bigg\{ &
	J \sum_{n,\sigma} \left( a_{n,\sigma} b_{n+1,\sigma} + a_{n+1,\sigma} b_{n,\sigma} \right) \nonumber                                                    \\
	                        & + \sum_{n,\sigma} V_n a_{n,\sigma} b_{n,\sigma} \nonumber  \\                                                                 
	                        & - (\alpha_z + \Delta) \sum_{n} \left( a_{n+1,\uparrow} b_{n,\downarrow} - a_{n,\uparrow} b_{n+1,\downarrow} \right) \nonumber \\
	                        & - (\alpha_z - \Delta) \sum_{n} \left( a_{n,\downarrow} b_{n+1,\uparrow} - a_{n+1,\downarrow} b_{n,\uparrow} \right)
	\bigg\}.
	\label{eq:hamiltonian_majorana}
\end{align}

We then follow the approach used in Ref.~\cite{degottardi2011topological} to construct the transfer matrix where the Heisenberg representation and the time-dependent Majorana modes $a_{k,\sigma}(t) = \alpha_{k,\sigma} e^{-i\omega_k t}$ and $b_{k,\sigma}(t) = \beta_{k,\sigma} e^{i\omega_k t}$  have been used. Furthermore,

\begin{align}
	\frac{1}{i}  [a_{k,\sigma},H] = \frac{d}{dt} a_{k,\sigma}, \text{and} \\
	\frac{1}{i}  [b_{k,\sigma},H] = \frac{d}{dt} b_{k,\sigma}.~~~~~~
\end{align}
To identify Majorana modes, we set $\omega = 0$ and get the following two coupled equations for $\beta$.
\small
\begin{align}
	J \beta_{k+1,\uparrow} + J \beta_{k-1,\uparrow} + V_k \beta_{k,\uparrow} - (\alpha_z + \Delta) \left(\beta_{k-1,\downarrow} - \beta_{k+1,\downarrow}\right) = 0,\\
	J \beta_{k+1,\downarrow} + J \beta_{k-1,\downarrow} + V_k \beta_{k,\downarrow} - (\alpha_z - \Delta) \left(\beta_{k+1,\uparrow} - \beta_{k-1,\uparrow}\right) = 0.~~~~~
\end{align}
\normalsize
A similar equation can be obtained for $\alpha$ whose transfer matrix would just be the inverse of the one found using $\beta$. Knowing the behavior of one completely determines the other. In addition, contrary to the spinless case, the terms mix the spin-up and spin-down components. Moreover, we can decouple these equations by applying the transformation:

$$U =
	\begin{pmatrix}
		\sqrt{\Delta+\alpha_z} & \sqrt{\Delta+\alpha_z}  \\
		\sqrt{\Delta-\alpha_z} & -\sqrt{\Delta-\alpha_z}
	\end{pmatrix}$$

The transformation gives two decoupled equations
\begin{equation}
	\begin{aligned}
		(J + \bar{\Delta}) \beta'_{k+1, \uparrow} + V_k \beta'_{k, \uparrow} + (J - \bar{\Delta}) \beta'_{k-1, \uparrow} = 0,\text{and}\\
		(J - \bar{\Delta}) \beta'_{k+1, \downarrow} + V_k \beta'_{k, \downarrow} + (J + \bar{\Delta}) \beta'_{k-1, \downarrow} = 0. ~~~~~
	\end{aligned}
	\label{decoupled_equations}
\end{equation}

where $$\begin{pmatrix}
		\beta'_{k+1, \uparrow} \\
		\beta'_{k+1, \downarrow}
	\end{pmatrix}
	=
	U^{-1}
	\begin{pmatrix}
		\beta_{k+1, \uparrow} \\
		\beta_{k+1, \downarrow}
	\end{pmatrix}.$$

 One should note that with RSOC, the up and down spins are not good quantum numbers and the $\beta'_{\uparrow}$ and $\beta'_{\downarrow}$ are just labels usually referred to as pseudo-spins. From the decoupled equations in Eq.~\eqref{decoupled_equations}, we identify two separate transfer matrices given by,

\begin{equation}
	\begin{pmatrix}
		\beta'_{k+1, \uparrow} \\
		\beta'_{k, \uparrow}
	\end{pmatrix} =
	\begin{pmatrix}
		-\frac{V_k}{J + \bar{\Delta}} & -\frac{J - \bar{\Delta}}{J + \bar{\Delta}} \\
		1                             & 0
	\end{pmatrix}
	\begin{pmatrix}
		\beta'_{k, \uparrow} \\
		\beta'_{k-1, \uparrow}
	\end{pmatrix}
	\label{eq:transfer_matrix_up}
\end{equation}

and

\begin{equation}
	\begin{pmatrix}
		\beta'_{k+1, \downarrow} \\
		\beta'_{k, \downarrow}
	\end{pmatrix} =
	\begin{pmatrix}
		-\frac{V_k}{J - \bar{\Delta}} & -\frac{J + \bar{\Delta}}{J - \bar{\Delta}} \\
		1                             & 0
	\end{pmatrix}
	\begin{pmatrix}
		\beta'_{k, \downarrow} \\
		\beta'_{k-1, \downarrow}
	\end{pmatrix}.
	\label{eq:transfer_matrix_down}
\end{equation}

It is important to note that these two transfer matrices for $\beta'_{\uparrow}$ and $\beta'_{\downarrow}$ can be used to calculate the Lyapunov exponent \cite{degottardi2013majorana,degottardi2011topological} as elaborately discussed in the next section.

\subsubsection{Estimation of the Lyapunov exponent}
\label{VB}
The Lyapunov exponent characterizes the localization properties of eigenstates and provides a powerful tool to identify topological phase transitions in quasiperiodic systems. Therefore, to find the topological phase boundaries with complex quasiperiodic potential as given above, we denote the transfer matrices derived in Eqs.~\eqref{eq:transfer_matrix_up} and \eqref{eq:transfer_matrix_down} as $T_{n,\uparrow}$ and $T_{n,\downarrow}$, respectively. Although we have two transfer matrices corresponding to the pseudo-spin states, we need only calculate the Lyapunov exponent using $T_{n,\uparrow}$ for our analysis. This choice is justified by the fundamental relationship between these matrices. In particular, the eigenvalues of $T_{n,\downarrow}$ are exactly the inverse of those of $T_{n,\uparrow}$.

It is well established that for a state to be localized, all eigenvalues of its transfer matrix must have magnitudes either all greater than one or all less than one. When one pseudo-spin state localizes to the right (eigenvalues $>1$), the other necessarily localizes to the left (eigenvalues $<1$). Therefore, determining the localization properties of one pseudo-spin component completely characterizes the localization behavior of the entire system, making the analysis of a single transfer matrix sufficient.

To obtain a closed analytic expression for the Lyapunov exponent, we follow the standard strategy used in Avila's global theory \cite{liu2021localization,10.1007/s11511-015-0128-7,10.1215/00127094-2017-0013,lin2021real}. 
In this approach, one studies how the Lyapunov exponent behaves when the phase of the quasiperiodic potential is shifted into the complex plane. and involves taking the limit $h \to \infty$ in
$
V_k = V\cos(2\pi \alpha k + ih) = V\cos(\phi + ih),
$
under which the potential simplifies to $V_k \to \tfrac{V}{2} e^{h} e^{-i\phi}$, and the transfer matrix reduces to
$$
	T_{k,\uparrow} \to     \begin{pmatrix}
		-\frac{V e^{h} e^{-i \phi}}{2(J + \bar{\Delta})} & 0 \\[6pt]
		0                                                & 0
	\end{pmatrix} + \mathcal{O}(1)$$

The transfer matrix for the entire system is then given by, 
\begin{equation}
	T(N) = T_{1,\uparrow} T_{2,\uparrow} \cdots T_{N,\uparrow}.
\end{equation}
The Lyapunov exponent can be estimated as $\gamma = \lim_{N \to \infty} \frac{1}{N} \log||T_{1,\uparrow} T_{2,\uparrow} \cdots T_{N,\uparrow}||$. Using these expressions, the final expression for the Lyapunov exponent becomes,

\begin{align}
	\gamma =
	\begin{cases}
		0,                                               & \text{if } V e^h < 2(J + \bar{\Delta}), \\
		\ln\left(\frac{V}{2(J+\bar{\Delta})}\right) + 
        h, & \text{otherwise.}
	\end{cases}
\end{align}

It is worth noting that for $\Delta = 0$, our model reduces to the non-Hermitian AAH model previously studied in Ref.~\cite{AditiAAH_PRB_2022}. In this case $\bar{\Delta}$ becomes purely imaginary, yielding $|J+\bar{\Delta}| = \sqrt{J^2 + \alpha_z^2}$, which reproduces the analytical result reported therein.


\begin{figure}[]
  \begin{tabular}{p{\linewidth}c}
      \centering
      \includegraphics[width=0.48\columnwidth, height=3.6cm]{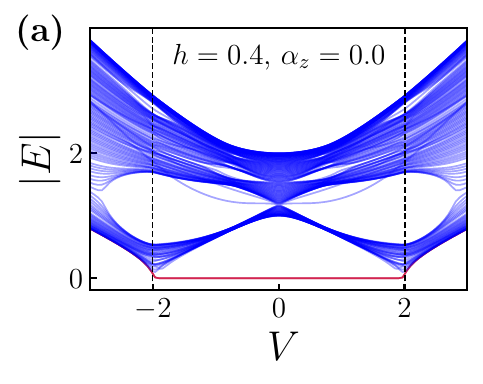}
      \includegraphics[width=0.48\columnwidth, height=3.6cm]{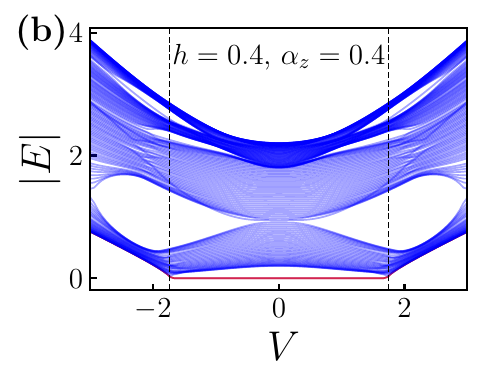}\\
      \caption{The absolute values of eigenenergies $|E|$ as a function of quasiperiodic potential strength $V$ calculated under OBC with the imaginary phase component $h$ fixed at $0.4$ with (a) $\alpha_z = 0.0$ (without RSOC) and (b) $\alpha_z=0.4$ (with RSOC). The red points indicate the presence of MZMs while the vertical dashed lines indicate the analytically predicted phase transition points at $|V| = 2e^{-h}(J+\bar{\Delta})$.}
      \label{fig:fig7}
  \end{tabular}
\end{figure}

\subsubsection{Numerical phase boundaries}
\label{VC}
To verify our analytical predictions, we perform exact diagonalization of the Hamiltonian and plot $|E|$ as a function of $V$ for different values of $\alpha_z$, similar to the previous case. Fig.~\ref{fig:fig7} demonstrates excellent agreement between our analytical prediction for the phase transition point and the numerical results. The closing of the energy gap, which signals the topological phase transition, occurs precisely at $V = 2e^{-h}(J+\bar{\Delta})$, as obtained analytically. Moreover, as $\alpha_z$ increases from 0 to 0.4 (comparing panels (a) and (b)), the topological transition point shifts to a lower value of $V$, confirming that RSOC reduces the span of the topological phase.

\begin{figure}[]
	\begin{tabular}{p{\linewidth}c}
		\centering
    \includegraphics[width=0.48\columnwidth, height=3.6cm]{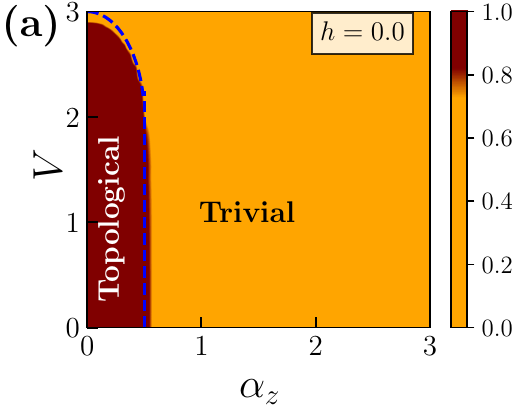}
    \includegraphics[width=0.48\columnwidth, height=3.6cm]{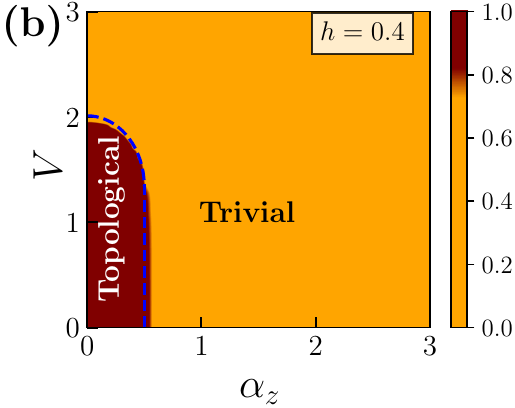}\\
		\caption{The real-space winding number of the system in the parameter space of the quasiperiodic potential strength $V$ and RSOC ($\alpha_z$). The maroon regime (W = 1) indicate topological phases, while the orange regime (W = 0) represents trivial phases. Panel (a) shows the phase diagram for $h=0.0$ (Hermitian case), while panel (b) shows $h=0.4$ (NH case). The non-Hermiticity reduces the topological region consistent with our analytical predictions.}
		\label{fig:fig8}
		\vspace{-0.2cm}
	\end{tabular}
\end{figure}

To further characterize the topological properties, we calculate the real-space winding number as done in the previous sections, which provides a direct measure of the system's topological invariant as discussed. Fig.~\ref{fig:fig8} reveals how the topological phase diagrams evolve with the NH parameter $h$. In the Hermitian case ($h=0.0$, panel (a)), the topological region is more extensive. As $h$ increases ($h=0.4$, panel (b)), the topological region shrinks. At the same strength $\alpha_z$ (which is low), the topological phase transition occurs at much lower value of $V$ when $h\neq 0$, as clearly illustrated. The boundary between topological and trivial phases matches our analytical prediction, i.e., $V = 2e^{-h}(J+\bar{\Delta})$.

\section{Conclusions}
\label{Conclusions}

In conclusions, in this work, we have systematically investigated the topological properties of spinful p-wave Kitaev chains with RSOC under two types of onsite potentials which bring in non-Hermiticity. Our analysis reveals several key findings.
First, in the Hermitian case, we demonstrated that the topological phase transition point remains fixed at $\mu = 2J$ when $\alpha_z < \Delta$, consistent with the original Kitaev model. However, when $\alpha_z > \Delta$, the system undergoes a qualitative change wherein the spectrum becomes gapless, and zero-energy edge modes merge into the bulk. This emergence of Fermi points occurs in pairs.
For the NH system with a gain/loss dictated on-site potential, we derived an analytical expression for the phase transition boundary, which agrees excellently with our numerical estimations. Crucially, unlike the Hermitian case, the topological region depends explicitly on the strength of $\alpha_z$, wherein the non-trivial phase shrinks with an increase in $\alpha_z$. In the quasiperiodic case, we employed the transfer matrix and Lyapunov exponent analysis to figure out the topological transition. We verify that the interplay of non-Hermiticity and RSOC can drastically alter the topological phase diagram, reducing the strength of $V$ for a given strength of $\alpha_z$. Overall, our findings advance our understanding of topological superconductivity in several important ways. First, they bridge the gap between idealized models and realistic materials by incorporating spin-orbit effects that are ubiquitous in experimental platforms. Second, they reveal the intricate interplay between topology and non-Hermiticity, which gives rise to several other interesting features in condensed matter physics.\\

\section*{Acknowledgments}\label{Sec:Acknowledgments}
A.C. thanks CSIR-HRDG (Govt. of India) for financial assistance via. File No-09/983(0047)/2020-EMR-I.

\appendix

\section{Derivation of the $k$-space Hamiltonian}\label{App_A}

Here, we derive the momentum-space representation of the Hamiltonian in Eq.~\eqref{eq:hamiltonian}.
We use the Fourier transform $c_{n,\sigma} = \frac{1}{\sqrt{N}} \sum_{k} e^{ikn} c_{k,\sigma}$
and introduce the Nambu spinor $\Psi_{k} = (c_{k,\uparrow}, c_{-k,\uparrow}^{\dagger}, c_{k,\downarrow}, c_{-k,\downarrow}^{\dagger})^{T}$ as already mentioned in Sec.~\ref{IIC}.

For clarity, we transform the terms one by one as given below:

\paragraph{Tight-binding term:}
\begin{align}
	& = J \sum_{n, \sigma}  \left( c_{n+1, \sigma}^\dagger c_{n, \sigma} + \text{H.c.} \right) \nonumber\\
	& = J \sum_{n, \sigma} \left( \frac{1}{\sqrt{N}} \sum_{k} e^{-ik(n+1)} c_{k, \sigma}^\dagger \right)
		\left( \frac{1}{\sqrt{N}} \sum_{k'} e^{ik'n} c_{k', \sigma} \right) + \text{H.c.} \nonumber \\
	& = \frac{J}{N} \sum_{n, \sigma} \sum_{k, k'} e^{i(k'-k)n} e^{-ik} c_{k, \sigma}^\dagger c_{k', \sigma} + \text{H.c.} \nonumber \\
	& = J \sum_{k, \sigma} e^{-ik} c_{k, \sigma}^\dagger c_{k, \sigma} + \text{H.c.} \nonumber                                       \\
	& = 2J \sum_{k, \sigma} \cos(k) c_{k, \sigma}^\dagger c_{k, \sigma}
\end{align}

\paragraph{p-wave pairing term:}
\small
\begin{align}
&= \Delta \sum_{n,\sigma,\sigma'}
\left( c_{n+1,\sigma}^{\dagger} c_{n,\sigma'}^{\dagger} + \text{H.c.} \right) \nonumber \\[4pt]
&= \frac{\Delta}{N} \sum_{n, k, k'} 
\Big[ e^{-i(k'+k)n} e^{-ik'} 
\big( c_{k', \uparrow}^\dagger c_{k, \downarrow}^\dagger 
     + c_{k', \downarrow}^\dagger c_{k, \uparrow}^\dagger \big) \Big]
+ \text{H.c.} \nonumber \\[4pt]
&= \Delta \sum_{k} e^{ik} 
      (c_{-k, \uparrow}^\dagger c_{k, \downarrow}^\dagger
   +
      c_{-k, \downarrow}^\dagger c_{k, \uparrow}^\dagger)
   + \text{H.c.} \nonumber \\[4pt]
&= \Delta \sum_{k} 
     \left( e^{ik} - e^{-ik} \right)
     c_{-k, \uparrow}^\dagger c_{k, \downarrow}^\dagger
   + \text{H.c.} \nonumber \\[4pt]
&= 2i\Delta \sum_{k} \sin k 
   \left(c_{-k, \uparrow}^\dagger c_{k, \downarrow}^\dagger 
        - c_{k, \downarrow} c_{-k, \uparrow} \right)
\end{align}

\paragraph{RSO interaction term:}
\small
\begin{align}
&= -\alpha_z \sum_{n} 
   \left( c_{n+1,\uparrow}^\dagger c_{n,\downarrow} 
        - c_{n+1,\downarrow}^\dagger c_{n,\uparrow} 
        + \text{H.c.} \right) \nonumber \\[4pt]
 &= \frac{-\alpha_z}{N} \sum_{n, k, k'} 
   e^{-i(k-k')n} e^{-ik} 
   \left( c_{k,\uparrow}^\dagger c_{k',\downarrow} 
        - c_{k,\downarrow}^\dagger c_{k',\uparrow} \right)
   + \text{H.c.} \nonumber \\[4pt]
&= -\alpha_z \sum_{k} e^{-ik} 
   \left( c_{k,\uparrow}^\dagger c_{k,\downarrow} 
        - c_{k,\downarrow}^\dagger c_{k,\uparrow} \right)
   + \text{H.c.} \nonumber \\[4pt]
&= 2i\alpha_z \sum_{k} \sin k 
   \left( c_{k,\uparrow}^\dagger c_{k,\downarrow} 
        - c_{k,\downarrow}^\dagger c_{k,\uparrow} \right)
\end{align}

\paragraph{On-site potential term:}
\begin{align}
	\sum_{n, \sigma} V'_n c_{n, \sigma}^\dagger c_{n, \sigma}
	 & = \frac{1}{N} \sum_{n, \sigma} \sum_{k, k'} V' c_{k, \sigma}^\dagger c_{k', \sigma} e^{i(k'-k)n} \nonumber \\
	 & = \sum_{k, \sigma} V' c_{k, \sigma}^\dagger c_{k, \sigma}
\end{align}

On combining all the terms, the Hamiltonian in k-space is given as,

\begin{equation}
	\begin{aligned}
		\mathcal{H}
		= \sum_{k} \Big[
			2J \cos(k) \sum_{\sigma} c_{k, \sigma}^\dagger c_{k, \sigma}
			+ 2i \Delta \sin(k) \left( c_{-k,\uparrow}^\dagger c_{k,\downarrow}^\dagger - c_{k,\downarrow} c_{-k,\uparrow} \right) \\
			\qquad\qquad
			+ 2i \alpha_z \sin(k) \left( c_{k,\uparrow}^\dagger c_{k,\downarrow} - c_{k,\downarrow}^\dagger c_{k,\uparrow} \right)
			+ V' \sum_{\sigma} c_{k, \sigma}^\dagger c_{k, \sigma}
		\Big].
	\end{aligned}
	\label{eq:bdg_hamiltonian_momentum_appendix}
\end{equation}

\section{The Hermitian p-wave superconductor with RSOC}\label{App:App_B}
\begin{figure}[ht!]
  \begin{tabular}{p{\linewidth}c}
      \centering
      \includegraphics[width=0.32\columnwidth, height=3.2cm]{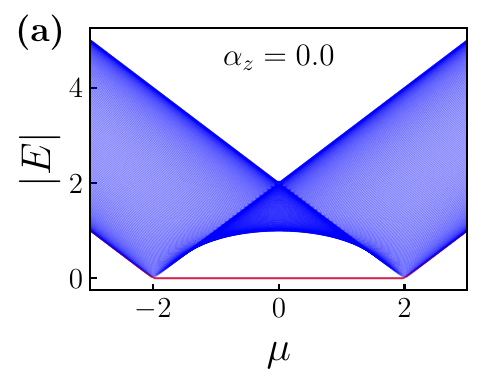}
      \includegraphics[width=0.32\columnwidth, height=3.2cm]{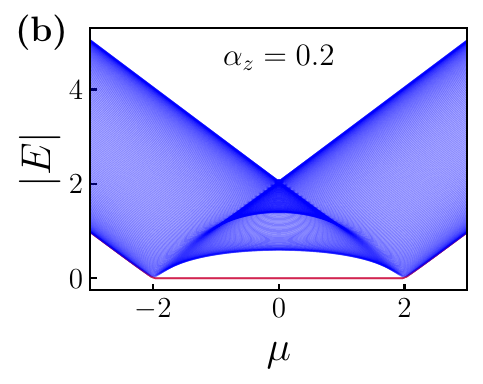}
      \includegraphics[width=0.32\columnwidth, height=3.2cm]{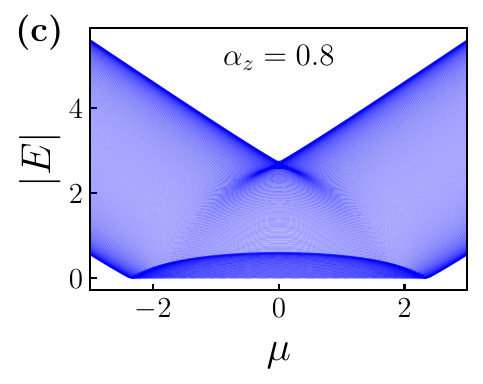}\\
      \caption{The absolute values of eigenenergies $|E|$ under OBC as a function of $\mu$ of the Hermitian potential ($\gamma=0.0$) at three distinct RSOC strengths: (a) $\alpha_z=0.0$ (absence of spin-orbit interaction), (b) $\alpha_z=0.2$ (intermediate coupling regime), and (c) $\alpha_z=0.8$ (strong coupling regime, i.e $\alpha_z>\Delta$). The red points indicate the presence of MZMs at the edges of the system. In the strong coupling regime the spectrum becomes gapless with the disappearance of the zero modes as is evident in (c).}
      \label{fig:fig1}
  \end{tabular}
\end{figure}

To appreciate the novelties in topological phases in a NH system, in this section, we first revisit the Hermitian version of the p-wave Kitaev chain with RSOC, where the on-site potential is given by $V_n' = \mu$. Since the topology is related to the energy spectrum of the Hamiltonian, we analyze the spectral behavior in Fig.~\ref{fig:fig1}.
To assess the role of $\alpha_z$ in the topology, we plot $|E|$ as a function of $\mu$ for different values of $\alpha_z$.
From Figs.~\ref{fig:fig1}(a), it is clear that when $\alpha_z=0$, MZMs are observed when $|\mu|<2J$. The MZMs disappear at $|\mu| = 2J$ which is concurrent to the topological transition from a non-trivial to trivial phase, and has been reported in the past literature \cite{Kitaev_2001}. Next, we consider a non-zero strength of the RSOC ($\alpha_z=0.2$) in Fig.~\ref{fig:fig1}(b) and demonstrate that the topological transition does not change with $\alpha_z$. Therefore, it is clear that RSOC does not affect the topology in a Hermitian p-wave superconductor as long as $\alpha_z<\Delta$. This can also be roughly understood from Eq.\ref{eq:phase_transition_non_hermitian}, since $\gamma=0$ when the system is Hermitian. It is important to note that as soon as $\alpha_z>\Delta$, the spectrum becomes gapless. In addition, the edge modes and the topological phase disappears and merges into the bulk of the system (Fig.~\ref{fig:fig1}(c).

\begin{figure}[ht!]
  \begin{tabular}{p{\linewidth}c}
      \centering
      \includegraphics[width=0.32\columnwidth, height=3.2cm]{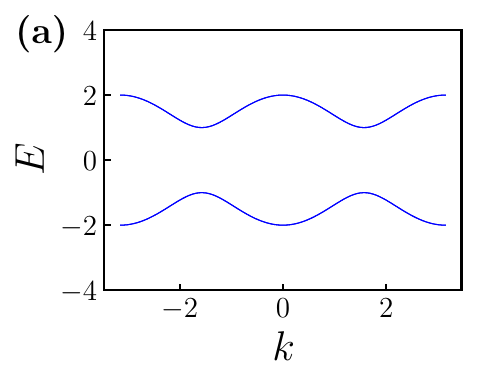}
      \includegraphics[width=0.32\columnwidth, height=3.2cm]{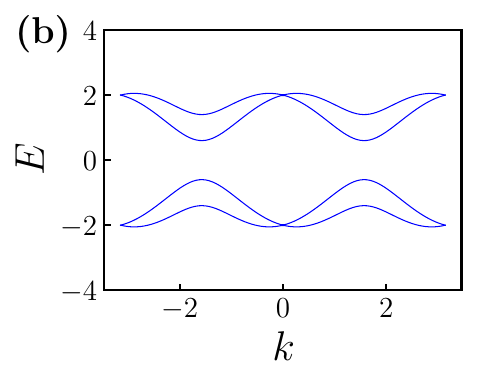}
      \includegraphics[width=0.32\columnwidth, height=3.2cm]{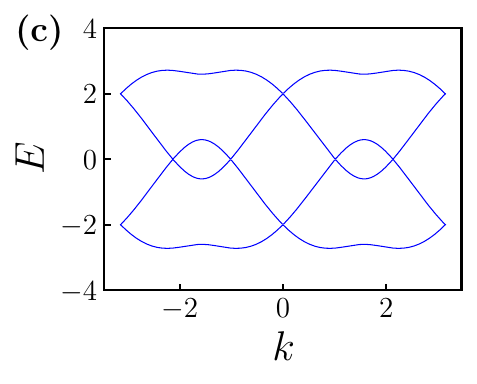}\\
      \caption{The energy bands for the Hermitian case as presented in Fig.~\ref{fig:fig1}. Panels illustrate the band structure at fixed chemical potential $\mu = 0.0$ with varying RSOC strengths: (a) $\alpha_z = 0.0$, corresponding to the conventional Kitaev model without spin-orbit coupling. (b) $\alpha_z = 0.2$, demonstrating the initial effects of moderate Rashba coupling, which induces band splitting and modifies the dispersion relation while preserving the overall band gap; and (c) $\alpha_z = 0.8$, in the strong coupling regime, characterized by band crossing and the disappearance of the band gap.}
      \label{fig:fig2}
  \end{tabular}
\end{figure}

\begin{figure}[ht!]
  \begin{tabular}{p{\linewidth}c}
      \centering
      \includegraphics[width=0.32\columnwidth, height=3.2cm]{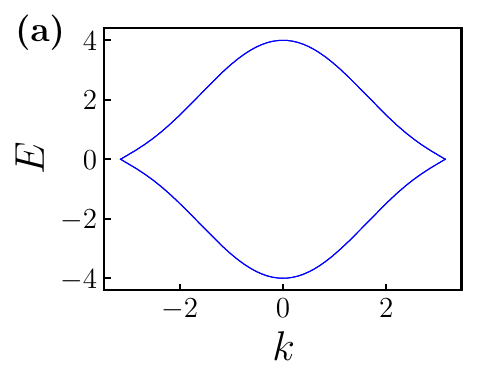}
      \includegraphics[width=0.32\columnwidth, height=3.2cm]{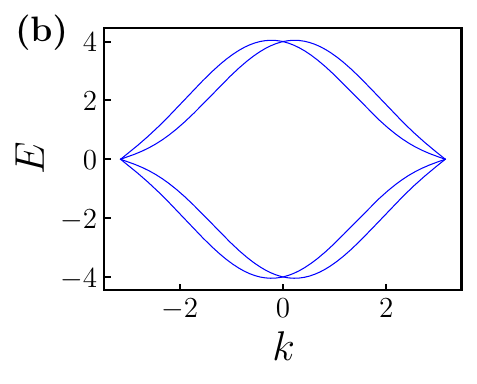}
      \includegraphics[width=0.32\columnwidth, height=3.2cm]{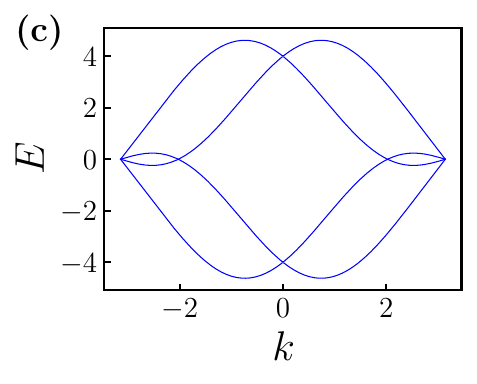}\\
      \caption{The energy bands plotted at the topological transition point $|\mu|=2J = 2$, where the system exhibits critical behavior. Panels display the evolution of band structure with increasing strength of RSOC: (a) $\alpha_z = 0.0$, depicting the conventional band gap closing at $k=\pi$. (b) $\alpha_z = 0.2$, showing how moderate spin-orbit coupling modifies the dispersion, splitting the bands; and (c) $\alpha_z = 0.8$, in the strong coupling regime, where in addition to band gap closing at $\pm\pi$, the bands cross at intermediate points.}
      \label{fig:figS2}
  \end{tabular}
\end{figure}

Furthermore, to obtain a clear picture of the energy bands in the k-space, we plot the dispersion relation using Eq.~\eqref{eq:bdg_hamiltonian_momentum_matrix} which is given by:
\tiny\begin{eqnarray}
  \epsilon_{1,\pm}(k) = \pm\sqrt{4 (\alpha_z^2 + \Delta^2) \sin^2(k) +
  \xi_k^2 + 2 \sqrt{(4 \alpha_z^2 \sin^2(k))(4 \Delta^2 \sin^2(k) + \xi_k^2})},~~~~~~~~~
\end{eqnarray}
\begin{eqnarray}
  \epsilon_{2,\pm}(k) = \pm\sqrt{4 (\alpha_z^2 + \Delta^2) \sin^2(k) +
  \xi_k^2 - 2 \sqrt{(4 \alpha_z^2 \sin^2(k))(4 \Delta^2 \sin^2(k) + \xi_k^2})}.~~~~~~~~~
\end{eqnarray}
 \normalsize Fig.~\ref{fig:fig2} demonstrates the band gap illustrating the insulating behavior in the absence of RSOC. Interestingly, we find the splitting of the energy bands in  Fig.~\ref{fig:fig2}(b) when the RSOC is turned on ($\alpha_z=0.2$), which is also expected from previous investigations. From Fig.~\ref{fig:fig2}(c), we find the band-crossing when $\alpha_z>\Delta$. We also demonstrate the behavior of the energy dispersion at the topological phase transition point $\mu = 2J$ in Fig.~\ref{fig:figS2}, where the energy bands touch, indicating the topological transition. In Fig.~\ref{fig:figS2}(a), corresponding to $\alpha_z = 0$, the two bands are fully degenerate throughout the Brillouin zone, reflecting the absence of RSOC-induced splitting. As $\alpha_z$ is increased to a finite value, shown in Fig.~\ref{fig:figS2}(b), the degeneracy is lifted and the bands separate, consistent with the expected RSOC-driven splitting of the spectrum. When the RSOC strength exceeds the pairing amplitude ($\alpha_z > \Delta$), the spectrum becomes gapless, as illustrated in Fig.~\ref{fig:figS2}(c) with the appearance of fermi points. 


\bibliography{references.bib}

\end{document}